\documentclass[12pt]{article}
\usepackage{array}
\usepackage{graphicx}
\usepackage{amssymb}
\usepackage{amsmath}
\usepackage{multirow}
\input{epsf}
\usepackage{cite}
\def\@fmsl@sh#1#2#3{\m@th\ooalign{$\hfil#1\mkern#2/\hfil$\crcr$#1#3$}}
 \def\eq#1\en{\begin{equation}#1\end{equation}}
\def\s[#1,#2]{[#1\stackrel{\star}{,}#2]}
\def\sx[#1,#2]{[#1\stackrel{\star_{x}}{,}#2]}

\newcommand{\nc}{\newcommand}
\nc{\beq}{\begin{equation}}
\nc{\eeq}{\end{equation}}
\nc{\beqa}{\begin{eqnarray}}
\nc{\eeqa}{\end{eqnarray}}

\def\bc{\begin{center}}
\def\ec{\end{center}}

\def\to{\rightarrow}

\def\gsim{\mathrel{\mathpalette\atversim>}}

\def\bc{\begin{center}}
\def\ec{\end{center}}

\def\gsim{\mathrel{\rlap{\lower4pt\hbox{\hskip1pt$\sim$}}

    \raise1pt\hbox{$>$}}}       

\def\gsim{\mathrel{\rlap{\lower4pt\hbox{\hskip1pt$\sim$}}
    \raise1pt\hbox{$>$}}}       

\begin{document}
\makeatletter
\def\fmslash{\@ifnextchar[{\fmsl@sh}{\fmsl@sh[0mu]}}
\def\fmsl@sh[#1]#2{%
  \mathchoice
    {\@fmsl@sh\displaystyle{#1}{#2}}%
    {\@fmsl@sh\textstyle{#1}{#2}}%
    {\@fmsl@sh\scriptstyle{#1}{#2}}%
    {\@fmsl@sh\scriptscriptstyle{#1}{#2}}}
\def\@fmsl@sh#1#2#3{\m@th\ooalign{$\hfil#1\mkern#2/\hfil$\crcr$#1#3$}}
\makeatother

\thispagestyle{empty}
\begin{titlepage}
\boldmath
\begin{center}
  \Large {\bf The Dual Standard Model and the 750 GeV Events at the LHC}
    \end{center}
\unboldmath
\vspace{0.2cm}
\begin{center}
{ {\large Xavier Calmet}\footnote{x.calmet@sussex.ac.uk}
}
 \end{center}
\begin{center}
{\sl Physics and Astronomy, 
University of Sussex,   Falmer, Brighton, BN1 9QH, UK 
}
\end{center}
\vspace{\fill}
\begin{abstract}
\noindent
The aim of this short paper is to discuss the recently observed excess at 750 GeV by both CMS and ATLAS in the light of the dual standard model. Within this framework it is natural to introduce neutral spin 0 and/or spin 2 $SU(2)$ glue mesons which could easily account for this observation if it is confirmed. The model predicts that these glue mesons would be part of $SU(2)$ triplets and that there must thus be charged counterparts of these glue mesons carrying a QED charge of $\pm 1$ with a spin 0 and/or 2 as well.
\end{abstract}  
\end{titlepage}



\newpage

The aim of this short paper is to discuss the recently observed excess at 750 GeV by both CMS \cite{CMS:2015dxe} and ATLAS \cite{ATLAS} in the light of the dual standard model \cite{Calmet:2000th,Calmet:2002mf}. This excess can be interpreted as a spin 0 or spin 2 resonance of mass  750 GeV  decaying into two photons. After briefly reviewing the dual standard model, we show that if the recent excess observed at CERN is a spin 0 or spin 2 resonance described within this framework, then this resonance must be the first sign of a new series of particles with similar masses. Indeed the spin 0 or 2 particle would be part of a $SU(2)$ triplet. Besides the neutral spin 0 or 2 particles,  one expects two states with the same spin and QED charges $\pm 1$. As we shall see, up to QED effects, these charged particles would have masses similar to that of their neutral counterpart.

The standard model provides a very successful description of the electroweak and strong interactions. It is based on the concept of local $SU(3)_c \times SU(2)_L \times U(1)_Y$ gauge invariance where $SU(3)_c$ is the gauge group that describes the strong interaction and $SU(2)_L$  and $U(1)_Y$ describe the weak and electromagnetic interactions. Besides the local gauge symmetries, there is another important but approximative global symmetry $SU(2)$ due to the structure of the Higgs potential of the standard model. This symmetry would be exact in the limit where the gauge coupling of the $U(1)_Y$ goes to zero. This custodial symmetry is particularly important when studying the standard model in its dual picture as we shall explain shortly. Indeed, it is well known that the standard model of particle physics can be defined in terms of gauge invariant fields. This formulation, while equivalent to the original presentation due to Glashow, Weinberg and Salam, is useful as it makes clear that particles are never isolated but always surrounded by virtual particles. Indeed a physical electron is never isolated from the photon and virtual pairs of particles constantly pop out of the vacuum. We can define fields which are gauge invariant under the local $SU(2)_L$ gauge symmetry. These ``physical" fields are given by \cite{Calmet:2000th,Calmet:2002mf,tHooft:1998pk,tHooft:1980xb,Visnjic:1987pj}
\begin{eqnarray}
  \Phi&=& \Omega^\dagger \phi,
  \\ \nonumber
\Psi^a_L&=&  \Omega^\dagger \psi^a_L
\\ \nonumber
W^i_\mu&=& \frac{i}{2g} \mbox{Tr} \left [ \ \Omega^\dagger \stackrel{\leftrightarrow}
{D}_\mu \Omega \tau^i \right ]
\\ \nonumber
({\cal D}_\mu)_{SU(2)_L}&=& \Omega^\dagger (D_\mu)_{SU(2)_L}
\Omega=\partial_\mu - i g W_\mu
\\ \nonumber
{\cal F}_{\mu \nu}&=&\frac{i}{g}[({\cal D}_\mu)_{SU(2)_L},({\cal D}_\nu)_{SU(2)_L}],
\end{eqnarray}
with $\phi^\dagger \stackrel{\leftrightarrow}
{D}_\mu \phi= \phi^\dagger D_\mu \phi - (D_\mu \phi)^\dagger \phi$ and where $\Omega$ is given by
\begin{eqnarray}
\Omega=\frac{1}{\sqrt{\phi^\dagger
    \phi}}\left(\begin{array}{cc}  \phi_2^* & \phi_1 
    \\ -\phi_1^* & \phi_2
  \end{array}
\right ).
\end{eqnarray}
Here $\Phi$ corresponds to the Higgs doublet, $\Psi^a_L$ to the left-handed fermions of the standard model (the index $a$ runs over all the lepton and quark flavors) and $W^i_\mu$ to the three electroweak bosons. Note that while these objects are gauge invariant under local $SU(2)_L$, they still transform under the global $SU(2)$ transformations:  $\Phi$ and $\Psi_L$ are global $SU(2)$ doublets while $W^i_\mu$ are triplets under the global $SU(2)$.  Using the unitarity gauge, which corresponds to the choice $\Omega=1$, one finds $\Phi\to \phi$, $\Psi
\to \psi$ and $W^i_\mu\to B^i_\mu$.  Note that we could generalize this to locally invariant fields under the full gauge group of the standard model \cite{Lavelle:1995ty,Calmet:2010cb}. 

It is easy to reformulate the standard model in terms of these new fields.  Starting from the standard model Lagrangian 
\begin{eqnarray}
{\cal L}&=&-\frac{1}{2} \mbox{Tr}F_{\mu \nu}F^{\mu \nu}-
\frac{1}{4} f_{\mu \nu}f^{\mu \nu}
+ i \bar \psi^a_L \left (
  (\fmslash D_\mu)_{SU(2)} - i\frac{1}{2} g' Y \fmslash {\cal A}
\right )  \psi^a_L
\\ \nonumber &&   
+  i \bar \psi^a_R ( \fmslash \partial - i\frac{1}{2} g' Y \fmslash {\cal A} )
\psi^a_R
\\ \nonumber &&
+ \left( \left (  (D_\mu)_{SU(2)} - i\frac{1}{2} g' Y {\cal A}_\mu \right)
  \phi \right)^\dagger
\left( \left (  (D^\mu)_{SU(2)} - i\frac{1}{2} g' Y {\cal A}^\mu \right)
  \phi \right)
\\ \nonumber &&
+ V(\phi^\dagger \phi) - G_u ( \bar \psi^a_L \phi \psi^a_R
+ \bar \psi^a_R \phi^\dagger \psi^a_L )    - G_d ( \bar \psi^a_L \bar \phi \psi^a_R + \bar \psi^a_R \bar \phi^\dagger \psi^a_L )
\end{eqnarray}
where $\bar \phi=i \tau_2 \phi^\star$, the fermion $\psi^a$ is a
generic fermion field and the covariant derivative is given by $({\cal
  D}_\mu)_{SU(2)}=\partial_\mu -i g B_\mu$, with $B_\mu=1/2 \tau^a
B^a_\mu$. We denote the Yukawa couplings by $G_u$ and $G_d$.  The
gauge dependent fields can be replaced by their $SU(2)_L$ gauge
invariant counterparts.  One obtains the dual standard model
\begin{eqnarray}
{\cal L}&=&-\frac{1}{2} \mbox{Tr} {\cal F}_{\mu \nu} {\cal F}^{\mu \nu}-
\frac{1}{4} f_{\mu \nu}f^{\mu \nu}
+ i \bar \Psi^a_L \left (
  (\fmslash {\cal D}_\mu)_{SU(2)_L} - i\frac{1}{2} g' Y \fmslash {\cal A}
\right )  \Psi^a_L
\\ \nonumber &&
+  i \bar \psi^a_R ( \fmslash \partial - i\frac{1}{2} g' Y \fmslash {\cal A} )
\psi^a_R
\\ \nonumber &&
+ \left( \left (  ({\cal D}_\mu)_{SU(2)_L} - i\frac{1}{2} g' Y {\cal A}_\mu \right)
  \Phi \right)^\dagger
\left( \left (  ({\cal D}^\mu)_{SU(2)_L} - i\frac{1}{2} g' Y {\cal A}^\mu \right)
  \Phi \right)
\\ \nonumber &&
+ V(\Phi^\dagger \Phi)- G_u ( \bar \Psi^a_L \Phi \psi^a_R
+ \bar \psi^a_R \Phi^\dagger \Psi^a_L )    - G_d ( \bar \Psi^a_L \bar \Phi \psi^a_R + \bar \psi^a_R \bar \Phi^\dagger \Psi^a_L ).
\end{eqnarray}
The scalar field potential is as usual
\begin{eqnarray}
  V(\phi^\dagger \phi)=\frac{1}{2}\lambda \left( \phi^\dagger \phi -\frac{1}{2}v^2 \right)^2,
\end{eqnarray}
its counterpart  in term of $SU(2)_L$ gauge invariant fields is given by
\begin{eqnarray}
  V(\Phi^\dagger \Phi)=\frac{1}{2}\lambda \left( \Phi^\dagger \Phi -\frac{1}{2}v^2 \right)^2.
\end{eqnarray}
This potential can be minimized if the field $\Phi$ is forced to form
the gauge invariant condensate
\begin{eqnarray}
\langle
\Phi^\dagger \Phi \rangle=\langle
\phi^\dagger \phi \rangle=\frac{1}{2} v^2.
\end{eqnarray}
As in the standard model, the gauge invariant charged vector bosons receive a mass term of the form $m_W=g v/2$, the fermions receive
masses of the type $m_u=G_u v/\sqrt{2}$ for the up-type fermions and
$m_d=G_d v/\sqrt{2}$ for the down-type fermions. We also see that a term
\begin{eqnarray}
\frac{1}{2} g'g{\cal A}_\mu W^3_\mu \Phi^\dagger \Phi
\end{eqnarray}
appears, which gives rise to a mixing between the $U(1)$ generator and
the $W^3$ gauge invariant field.  After diagonalization according to
\begin{eqnarray}
         A_\mu&=& \sin{\theta_W} W^3_\mu+ \cos{\theta_W}{\cal A}_\mu
         \nonumber \\
         Z_\mu&=& \cos{\theta_W} W^3_\mu-  \sin{\theta_W}{\cal A}_\mu,
\end{eqnarray}
we can identify the photon $A_\mu$ and the $Z_\mu$ boson which are identical to the photon and $Z$ boson of the standard model.

The dual standard model \cite{Calmet:2000th,Calmet:2002mf} is identical to the usual standard model and should not be confused with composite models of the electroweak interactions such as that of Abbott and Farhi \cite{AB1,AB2,Claudson:1986ch}.  While the dual formulation is physically equivalent to the standard model, it is a useful starting point to discuss potential physics effects beyond the standard model. In particular, if one takes the dual picture seriously there must be a relation between the Higgs boson's mass and that of the three electroweak bosons, using this duality we were able to estimate the Higgs boson's mass and obtained a prediction of 129.6 GeV \cite{Calmet:2001rp} which is not far off from the measured Higgs boson's mass at 125 GeV. 

In the dual picture of the standard model, it is possible to construct more gauge invariant fields than those corresponding to the standard model field content. For example, we can construct objects which are a superposition of gauge bosons which we shall call electroweak glue mesons\footnote{A justification for this name is that the operators given in Eqs. (\ref{1}) and (\ref{2}) correspond in the unitary gauge respectively to $F^{\mu\nu}F^{\alpha\beta} \epsilon_{\mu\nu\alpha\beta}$ and $F_\mu^\alpha F_{\alpha\nu} -1/4 g_{\mu\nu}  F^{\alpha\beta} F_{\alpha\beta}$ which have been used to represent true glue mesons in QCD \cite{Fritzsch:1975tx}.}  despite the fact that this is not a composite model. For example, since there are three gauge bosons in $SU(2)_L$, we can construct ``glue mesons'' containing two gauge bosons
\begin{eqnarray} \label{1}
S^{ij}&=& \frac{-	1}{4g^2} \mbox{Tr} \left [ \ \Omega^\dagger \stackrel{\leftrightarrow}
{D}_\mu   \Omega \tau^i   \Omega^\dagger  \stackrel{\leftrightarrow}{D}^\mu \Omega \tau^j  \right ]\to B_\mu^i B^{j \mu}
\\ \label{2}
 T^{ij}_{\mu\nu}&=& \frac{-	1}{4g^2} \mbox{Tr}  \left [ \ \Omega^\dagger \stackrel{\leftrightarrow}
{D}_{ \{ \mu}   \Omega \tau^i \Omega^\dagger  \stackrel{\leftrightarrow}{D}_{\nu \}} \Omega \tau^j \right ] \to B_\mu^i B_\nu^j+B_\nu^i B_\mu^j
\end{eqnarray}
or in matrix form (using the standard notation $W^\pm_\mu=(B^1_\mu \mp i B^2_\mu)/\sqrt{2})$:
\begin{eqnarray}
S&=&
\tau^iS^{ij}\tau^j= 
\left(\begin{array}{cc}  
B_\mu^3 B^{3 \mu}  + 2 W^+_\mu W^{- \mu} &
 \sqrt{2}  B_\mu^3 W^{+ \mu}  - \sqrt{2} W^+_\mu B^{3 \mu}
 \\
 \sqrt{2} W^-_\mu B^{3 \mu}  -  \sqrt{2} B^3_\mu W^{- \mu} &
B_\mu^3 B^{3 \mu}  + 2 W^+_\mu W^{- \mu}   
  \end{array}
\right ).
\\
T_{\mu\nu}&=&
\tau^iT^{ij}_{\mu\nu}\tau^j= 
\left(\begin{array}{cc}  
B_{ \{ \mu}^3 B^3_{ \nu \} }  + 2 W^+_{ \{ \mu} W^-_{\nu \} }  &
 \sqrt{2}  B_{ \{ \mu}^3 W^+_{\nu \} }   - \sqrt{2} W^+_{ \{ \mu} B^3_{\nu \} } 
 \\
 \sqrt{2} W^-_{ \{ \mu} B^3_{\nu \} }   -  \sqrt{2} B^3_{ \{ \mu} W^-_{\nu \} }  &
B_{ \{ \mu}^3 B^3_{ \nu \} }   + 2 W^+_{ \{ \mu} W^-_{\nu \} }    
  \end{array}
\right ).
\end{eqnarray}
There are thus three spin 0 states with QED charges 0 and $\pm1$ and three spin 2 states (d-waves \cite{Calmet:2001yd}
) with QED charges 0 and $\pm 1$. The charge 0 states are a superposition two electroweak boson $Z Z$ or $W^+W^-$ and the states carrying charges $\pm 1$ are of the form $Z W^+$ or $Z W^-$. It is obviously possible to define states built from more than two gauge bosons as well as states involving left-handed fermions 
\begin{eqnarray}
\Psi^{a\star}_L&=& \frac{i}{2g}  \Omega^\dagger  \fmslash{D}  \psi^a_L\to  \fmslash{B}^i \tau^i \psi_L^a. 
\end{eqnarray}
Note that due to the chiral nature of the interactions of the standard model, it is not straightforward to write a mass term for the excited fermions in the effective Lagrangian. We could simply introduce a right-handed singlet $\psi^{a\star}_R$ in analogy to the fermions of the standard model. We will not investigate this question further here, but shall simply assume that if these fermions exist, they are very massive.

Because there are no states corresponding to $S^{ij}$ and $T^{ij}_{\mu\nu}$ in the standard model, we do not know how to couple these states to the particles of the standard model. The only guiding principle at our disposal is that we must preserve the approximate global $SU(2)$ which is only broken by the $U(1)_Y$ interaction. However, in the limit in which $U(1)_Y$ decouples from the $SU(2)_L$ sector, we must recover this symmetry.   The new fermions $\Psi^{a\star}_L$ that are charged under $SU(3)$ will behave as heavy quarks, they can also couple to the usual weak bosons of the standard model. They carry the hypercharge of their standard model counterparts and will thus couple to the hyperphoton. These new fermions could generate couplings between the gluons, the photon and electroweak bosons with the new scalar fields and the tensor modes which can be parametrized using effective field theory techniques. The coupling of the neutral scalar field to the spin 1 bosons of the standard model is given by
\begin{eqnarray}
L_{spin 0}=  S^0\sum_{a=1}^3 \frac{c_a}{\Lambda} \left (\frac{1}{4} {}^a\!F_{\alpha \beta} \  {}^a\!F^{\alpha \beta}\right) 
\end{eqnarray}
where  $S^0$ is the neutral spin 0 state and $ {}^a\!F_{\alpha \beta}$ are the field strength tensors of the $U(1)_Y$, $SU(2)_L$ and $SU(3)_c$ gauge fields. In the basis of the gauge boson mass eigenstates one finds:
\begin{eqnarray}
L_{spin 0}&=& \frac{S^0}{\Lambda} \left ( \frac{ c^{(0)}_{\gamma \gamma}}{4}  A_{\alpha \beta}A^{\alpha \beta}  + \frac{c^{(0)}_{\gamma Z} }{4}  A_{\alpha \beta}Z^{\alpha \beta} 
+  \frac{c^{(0)}_{Z Z}}{4}  Z_{\alpha \beta}Z^{\alpha \beta} + \frac{c^{(0)}_{W W}}{4} \eta^{\mu\nu}  W_{\alpha \beta}W^{\alpha \beta} 
 \right .  \nonumber
 \\ 
 &&
 \left . 
+ \frac{c^{(0)}_{gg}}{4}   G_{\alpha \beta}G^{\alpha \beta}    \right)
\end{eqnarray}
with 
\begin{eqnarray}
c^{(0)}_{\gamma \gamma} &=& c^{(0)}_1 \cos^2\theta_W +c^{(0)}_2 \sin^2\theta_W \\
c^{(0)}_{\gamma Z} &=&(c^{(0)}_1 -c^{(0)}_2)  \sin 2\theta_W  \\
c^{(0)}_{Z Z} &=& c^{(0)}_1 \sin^2\theta_W +c^{(0)}_2 \cos^2\theta_W \\
c^{(0)}_{W W}&=&2 c^{(0)}_2\\
c^{(0)}_{gg}&=& c^{(0)}_3.
\end{eqnarray}
The coefficients $c^{(0)}_i$ are free parameters.

For the spin 2 tensor, we have the following Lagrangian
\begin{eqnarray}
L_{spin 2}=  T^0_{\mu\nu} \sum_{a=1}^3 \frac{c^{(2)}_a}{\Lambda} \left (\frac{1}{4} \eta^{\mu\nu} \  {}^a\!F_{\alpha \beta} \  {}^a\!F^{\alpha \beta}  - 
{}^a\!F_{\alpha}^\mu \  {}^a\!F^{\alpha \nu} \right) 
\end{eqnarray}
where  $T^0_{\mu\nu}$ is the neutral spin 2 d-wave and as before $ {}^a\!F_{\alpha \beta}$ are the field strength tensors of the $U(1)_Y$, $SU(2)_L$ and $SU(3)_c$ gauge fields. In the basis of the gauge boson mass eigenstates one finds:
\begin{eqnarray}
L_{spin 2}&=& \frac{T^0_{\mu\nu}}{\Lambda} \left ( c^{(2)}_{\gamma \gamma} \left (\frac{1}{4} \eta^{\mu\nu}  A_{\alpha \beta}A^{\alpha \beta}  - 
A_{\ \alpha}^\mu A^{\alpha \nu} \right) + c^{(2)}_{\gamma Z} \left (\frac{1}{4} \eta^{\mu\nu}  A_{\alpha \beta}Z^{\alpha \beta}  - 
A_{\ \alpha}^\mu Z^{\alpha \nu} \right) \nonumber
 \right . \\  \nonumber
&& \left .
+ c^{(2)}_{Z Z} \left (\frac{1}{4} \eta^{\mu\nu}  Z_{\alpha \beta}Z^{\alpha \beta}  - 
Z_{\ \alpha}^\mu Z^{\alpha \nu} \right) + c^{(2)}_{W W} \left (\frac{1}{4} \eta^{\mu\nu}  W_{\alpha \beta}Z^{\alpha \beta}  - 
A_{\ \alpha}^\mu Z^{\alpha \nu} \right) 
 \right . 
 \\ 
 &&
 \left . 
+ c^{(2)}_{gg} \left (\frac{1}{4} \eta^{\mu\nu}  G_{\alpha \beta}G^{\alpha \beta}  - 
G_{\ \alpha}^\mu G^{\alpha \nu} \right)  \right )
\end{eqnarray}
with 
\begin{eqnarray}
c^{(2)}_{\gamma \gamma} &=& c^{(2)}_1 \cos^2\theta_W +c^{(2)}_2 \sin^2\theta_W \\
c^{(2)}_{\gamma Z} &=&(c^{(2)}_1 -c^{(2)}_2)  \sin 2\theta_W  \\
c^{(2)}_{Z Z} &=& c^{(2)}_1 \sin^2\theta_W +c^{(2)}_2 \cos^2\theta_W \\
c^{(2)}_{W W}&=&2 c^{(2)}_2\\
c^{(2)}_{g g}&=& c^{(2)}_3.
\end{eqnarray}
The coefficients $c^{(2)}_i$ are free parameters.

The neutral spin 0 or spin 2 glue mesons could easily account for the recent events observed at CERN corresponding to a 750 GeV particle decaying to two photons. Let us first consider the neutral scalar field $S^0$, using the narrow width approximation one finds (see e.g. \cite{Djouadi:2005gi}
):
 \begin{eqnarray} 
 \sigma(\mbox{proton} \ \mbox{proton} \to S^0 \to \gamma \gamma)=\sigma(\mbox{proton} \ \mbox{proton} \to S^0) \mbox{Br}(  S^0 \to \gamma \gamma).
 \end{eqnarray}
The production cross section at the parton level is given by 
\begin{eqnarray}
\hat \sigma(g g\to S^0)=\frac{(c^{(0)}_{gg})^2 \pi}{4} \frac{m^2_{S^0}}{\Lambda^2} \delta(\hat s-m^2_{S^0}).
\end{eqnarray}

Using the results of \cite{Low:2011gn}, it is straightforward to calculate the decay width of the neutral glue scalar meson to the electroweak bosons and gluons of the standard model:
\begin{eqnarray}
    \Gamma(S^0\to\gamma\gamma)&=&\frac{(c^{(0)}_{\gamma\gamma})^2}{4 \pi} \frac{m^3_{S^0}}{\Lambda^2} \\
    \Gamma(S^0\to g g)&=&2 \frac{(c^{(0)}_{g g})^2}{\pi} \frac{m^3_{S^0}}{\Lambda^2}\\
     \Gamma(S^0\to \gamma Z)&=&\frac{1}{8} \frac{(c^{(0)}_{\gamma Z})^2}{\pi} \frac{m^3_{S^0}}{\Lambda^2} \left ( 1 -\frac{m^2_Z}{m^2_{S^0}} \right )^3\\
    \Gamma(S^0\to W W)&=&64 \pi^3  (c^{(0)}_{W W})^2 \frac{m^5_{S^0}}{\Lambda^2 m_W^2} \sqrt{1- 4 \frac{m^2_W}{m^2_{S^0}}} \left(1-4 \frac{m^2_W}{m^2_{S^0}} + 12 \frac{m^4_W}{m^4_{S^0}} \right)\\
     \Gamma(S^0\to W W)&=&32 \pi^3  (c^{(0)}_{Z Z})^2 \frac{m^5_{S^0}}{\Lambda^2 m_Z^2} \sqrt{1- 4 \frac{m^2_Z}{m^2_{S^0}}} \left(1-4 \frac{m^2_Z}{m^2_{S^0}} + 12 \frac{m^4_Z}{m^4_{S^0}} \right).
\end{eqnarray}

Similarly, for the neutral d-wave could be produced as a resonance and the narrow width approximation can be used to calculate the cross section $\sigma(\mbox{proton} \ \mbox{proton} \to T^0 \to \gamma \gamma)$. The widths can easily be calculated.  One has \cite{Han:1998sg}:
\begin{eqnarray}
\Gamma(T^0_{\mu\nu} \to \gamma \gamma) &=& \frac{(c^{(2)}_{\gamma\gamma})^2 m^3_{T^0_{\mu\nu}}}{80 \pi \Lambda^2}\\
\Gamma(T^0_{\mu\nu} \to Z \gamma) &=& \frac{(c^{(2)}_{\gamma Z})^2 m^3_{T^0_{\mu\nu}}}{160 \pi \Lambda^2} \left ( 1 -\frac{m^2_Z}{m^2_{T^0_{\mu\nu}}} \right )^3 \left ( 1 + \frac{m^2_Z}{2 m^2_{T^0_{\mu\nu}}} + \frac{m^4_Z}{6 m^4_{T^0_{\mu\nu}}} \right ) \\
\Gamma(T^0_{\mu\nu} \to Z Z) &=& \frac{(c^{(2)}_{ZZ})^2 m^3_{T^0_{\mu\nu}}}{80 \pi \Lambda^2} \sqrt{  1 - 4\frac{m^2_Z}{m^2_{T^0_{\mu\nu}}} }\left ( 1 - 3 \frac{m^2_Z}{ m^2_{T^0_{\mu\nu}}} + 6  \frac{m^4_Z}{m^4_{T^0_{\mu\nu}}} \right ) \\
\Gamma(T^0_{\mu\nu} \to W W) &=& \frac{(c^{(2)}_{WW})^2 m^3_{T^0_{\mu\nu}}}{160 \pi \Lambda^2} \sqrt{  1 - 4\frac{m^2_W}{m^2_{T^0_{\mu\nu}}} }\left ( 1 - 3 \frac{m^2_W}{ m^2_{T^0_{\mu\nu}}} + 6  \frac{m^4_W}{m^4_{T^0_{\mu\nu}}} \right ) \\
\Gamma(T^0_{\mu\nu} \to gg) &=& \frac{(c^{(2)}_{gg})^2 m^3_{T^0_{\mu\nu}}}{10 \pi \Lambda^2}
\end{eqnarray}

Clearly, the 750 GeV excess could easily be explained by either the production of the spin 0 or spin 2 glue mesons discussed above. The analysis performed in e.g. \cite{Ellis:2015oso,Gupta:2015zzs,Han:2015cty,Sanz:2016auj} of the 750 GeV resonance as a spin 0 or spin 2 particle applies to our model. Because of the global $SU(2)$ symmetry, we know that besides the neutral spin 0 or spin 2 particles, we are expecting a charged counterpart with a mass close to 750 GeV modulo some small quantum electrodynamics effects in analogy to the mass difference between the charge electroweak bosons and the Z-bosons. The charged glue mesons are however more difficult to produce than their neutral counterparts as this would have happen from the fusion of electroweak bosons. Also their decays principally in charged fermions via $S/T \to W+Z \to 4 \ \mbox{fermions}$ are less straightforward than the clear $\gamma \gamma$ signature of decay of a neutral glue meson.

 If the 750 GeV resonance is confirmed by future measurements at the LHC, the dual standard model and its extension as described above is a very natural framework to account for this effect. Within that model one expects for three spin 0 and spin 2 glue mesons, there are charged $SU(2)$ partners corresponding to the neutral resonance which may have been recently observed. Clearly if this picture is correct, higher spin resonances in $SU(2)$ multiplets are also expected as well as more exotic particles corresponding to "excitations" of the known Higgs boson and fermions. While the Lagrangian proposed above is not renormalizable in the usual sense, it could be the first sign that asymptotic safety \cite{Calmet:2010ze} is truly relevant to the electroweak interactions.

{\it Acknowledgments:}
This work is supported in part by the Science and Technology Facilities Council (grant number  ST/L000504/1).
 

\bigskip


\end{document}